\pdfoutput=1
\documentclass[11pt]{article}
\usepackage[margin=2.6cm]{geometry}
\usepackage[english]{babel} 
\usepackage[utf8]{inputenc}
\usepackage{indentfirst}
\usepackage{amsmath}
\usepackage{amssymb}
\usepackage{url}
\usepackage{graphicx}
\usepackage{braket}
\usepackage{slashed}
\usepackage{cancel}
\usepackage[font=footnotesize,labelfont=bf]{caption}
\usepackage[usenames, dvipsnames]{xcolor}
\usepackage{booktabs}
\usepackage{dsfont}
\usepackage{pdflscape}
\usepackage{float}
\usepackage{soul}

\usepackage[colorlinks=true, urlcolor=MidnightBlue, linkcolor=MidnightBlue, citecolor=Mahogany, pdfborder={0 0 0}]{hyperref}
\usepackage[sort&compress,numbers]{natbib}
\usepackage{doi}
\usepackage[capitalise]{cleveref}
\newcommand{\amu}{$a_\mu^{\textrm{HLbL}}$}
\newcommand{\amuA}{$a_\mu^{\textrm{HLbL};A}$}

\title{\Large\bfseries{Axial-vector exchange contribution to the hadronic light-by-light piece of the muon anomalous magnetic moment}}
\author{\normalsize{
        Pablo Roig{\color{Mahogany}\thanks{proig@fis.cinvestav.mx}}$^{\color{Mahogany}{\ a}}$
        Pablo Sanchez-Puertas{\color{Mahogany}\thanks{psanchez@ifae.es}}$^{\color{Mahogany}{\ b}}$,
        } \vspace{0.2cm}  \\ 
        {\small{$^{\color{Mahogany}{a}}$\textit{Centro de Investigaci\'on y de Estudios Avanzados del IPN (Cinvestav),}}}\\
        {\small{\textit{Apdo. Postal 14-740, 07000 Ciudad de M\'exico, M\'exico}}} \\
        {\small{$^{\color{Mahogany}{b}}$\textit{Institut de F{\'i}sica d'Altes Energies (IFAE),}}} \\
        {\small{\textit{The Barcelona Institute of Science and Technology (BIST)}}}, \\
        {\small{\textit{Campus UAB, E-08193 Bellaterra (Barcelona), Spain}}} \\
}
\date{}

\begin{document}\renewcommand{\abstractname}{\vspace{-\baselineskip}} \maketitle
\begin{abstract} 
In this work we study the axial contributions to the hadronic light-by-light piece of the muon anomalous magnetic moment. We point out some theoretical ambiguities in previous estimates, and opt to perform a new evaluation using resonace chiral theory, that is free of them. As a result, we obtain $a_{\mu}^{\textrm{HLbL};A} = \left(0.8^{+3.5}_{-0.8}\right)\cdot 10^{-11}$, that might suggest a smaller value than most recent calculations, underlining the relevance of the off-shell prescription and the need for future work along this direction. Further, we find that our results depend critically on the asymptotic behavior of the form factors, and as such, emphasizes the relevance of future experiments at large photon virtualities. In addition, we present general results regarding the involved axial form factors description, comprehensively examining (and relating) the current approaches, that shall be of general interest.
\end{abstract}

\section{Introduction}
\subsection{Overview of the muon g-2: the importance of hadronic contributions}
The anomalous magnetic moment of a charged lepton $\ell$, $a_\ell=(g_\ell-2)/2$ is nonvanishing because of quantum radiative corrections, and has played a key role since its first measurement showing its nonvanishing value \cite{Foley:1948zz, Kusch:1948mvb} for $\ell=e$, confirmed immediately after with the famous Schwinger computation of $a_\ell=\alpha/(2\pi)+\mathcal{O}(\alpha^2)$ \cite{Schwinger:1948iu}. Over the years, it has been (and it still is) one of the most stringent tests of the whole Standard Model, thanks to the increasing accuracy of its determination over time ($\ell=e,\mu$) and the improved theoretical computations with reduced uncertainties that became available. This makes it an extremely sensitive probe of new physics that, if heavy, would naively shift $a_\ell$ with the scaling $m_\ell^2/M^2$ ($M$ being the heavy new physics mass). This explains why, despite $a_e$ is measured $2400$ times more precisely than $a_\mu$, the latter is still more sensitive to heavy new particles than $a_e$ by a factor $\sim18$.\footnote{There are exceptions to this counting, see Ref.~\cite{Giudice:2012ms}. Also, there are proposals to measure $a_e$ to a precision high enough as to compete with the future $a_{\mu}$ experiments~\cite{Terranova:2013vfa}. It is nevertheless extremely interesting that, while the measurement of $a_e$ \cite{Hanneke:2008tm} agrees with the former prediction in \cite{Aoyama:2017uqe} (that uses as an input the previous values for $\alpha$ from Refs.~\cite{Mohr:2015ccw,Mohr:2018hvt}), it is in tension, at the 2.5 $\sigma$ level, when employing the most recent and precise determination of $\alpha$~\cite{Parker:2018vye}. Note at this respect that, as opposed to $a_{\mu}$, the $a_{e}$ uncertainty is dominated by that of $\alpha$~\cite{Aoyama:2017uqe}.}
The latest measurements of $a_\mu$ \cite{Bennett:2002jb, Bennett:2004pv,  Bennett:2006fi} yield \cite{Tanabashi:2018oca}
\begin{equation}\label{measurement}
    a_\mu^{\mathrm{exp}}\,=\,(116592091\pm63)\cdot10^{-11}\,,
\end{equation}
while the weighted average of the most recent evaluations \cite{Jegerlehner:2017gek,Keshavarzi:2018mgv, Davier:2019can} of the SM contributions reads\footnote{This arises from $a_\mu^{\mathrm{SM}}\,=\,1.16591783(35)$~\cite{Jegerlehner:2017gek}, $a_\mu^{\mathrm{SM}}\,=\,1.16591820.4(35.6)$~\cite{Keshavarzi:2018mgv}, $a_\mu^{\mathrm{SM}}\,=\,1.16591830(48)$~\cite{Jegerlehner:2017gek}.}
\begin{equation}\label{prediction}
    a_\mu^{\mathrm{SM}}\,=\,(116591807\pm38)\cdot10^{-11}\,,
\end{equation}
showing a tantalizing $3.9\sigma$ discrepancy with respect to the measurement (\ref{measurement}). This has motivated two further experiments: one at FNAL, aiming to achieve an error around $16\cdot 10^{-11}$~\cite{Grange:2015fou}, and a second one at J-PARC, aiming for an error around $50\cdot 10^{-11}$~\cite{Abe:2019thb}.

The amazing precision of the theoretical determination in \cref{prediction} is possible thanks to the complete $\mathcal{O}(\alpha^5)$ computation of the QED contributions \cite{Aoyama:2012wk,  Aoyama:2014sxa, Aoyama:2017uqe} and of the electroweak contributions to two loops (including the  leading logarithms  from an additional loop) \cite{Gnendiger:2013pva,Czarnecki:2002nt,Knecht:2002hr,Ishikawa:2018rlv}, which warrant an associated uncertainty at the level of $\lesssim 1\times10^{-11}$. Still, the error of~\cref{measurement} is a factor $\sim40$ larger, because of the uncertainties associated to the hadronic contributions \cite{Prades:2009tw, Jegerlehner:2009ry, Benayoun:2014tra}, as we will discuss next.

There are two main types of hadronic contributions to $a_\mu$: the so-called hadronic vacuum polarization (HVP) and the hadronic light-by-light (HLbL) scattering, which are $\mathcal{O}(\alpha^2)$ and $\mathcal{O}(\alpha^3)$, respectively. The $38\cdot10^{-11}$ uncertainty in the SM prediction of $a_\mu$ above comes from their leading order components (at the next-to-leading order they are known \cite{Kurz:2014wya, Colangelo:2014qya} precisely enough). Despite their dominantly nonperturbative  nature, it has long been known \cite{Bouchiat:1961lbg, Brodsky:1967sr} how to obtain a data-driven extraction of the LO HVP contribution via dispersion relations, that provide an immediate connection to the $e^+e^-\to\textrm{hadrons}$ cross section. Although the resulting error used to dominate the total uncertainty in \cref{prediction}, relegating the HLbL to a second place, the successive improvements on the HVP side (with current errors around  $35\cdot10^{-11}$~\cite{Jegerlehner:2017gek,Keshavarzi:2018mgv, Davier:2019can}) demanded a dedicated theory effort for the HLbL piece (with former errors around $30\times 10^{-11}$~\cite{Jegerlehner:2005fs,Prades:2009tw}). More important, to fully benefit from the future measurements of $a_\mu$ at FNAL and J-PARC, a reduction of errors at around $10^{-10}$ is required.

For a long time, the leading order HLbL could not be computed in a data-driven way and it was difficult to evaluate the  model-dependence associated to it \cite{deRafael:1993za, Bijnens:1995cc, Bijnens:1995xf, Bijnens:2001cq, Hayakawa:1995ps, Hayakawa:1996ki, Hayakawa:1997rq, Knecht:2001qg, Knecht:2001qf, Blokland:2001pb, RamseyMusolf:2002cy, Melnikov:2003xd, Kampf:2011ty, Engel:2012xb, Masjuan:2012wy, Engel:2013kda, Roig:2014uja, Bijnens:2016hgx}. Recently, there has been a tremendous effort in this direction \cite{Colangelo:2014dfa, Colangelo:2014pva, Colangelo:2015ama, Colangelo:2017qdm, Colangelo:2017fiz}
 yielding precise numerical results for the two-pion  \cite{Colangelo:2017qdm, Colangelo:2017fiz}, one-pion  \cite{Masjuan:2017tvw, Hoferichter:2018dmo, Hoferichter:2018kwz}, and $\eta,\eta'$~\cite{Masjuan:2017tvw} contributions---that are the most relevant ones.\footnote{Remarkable progress in the evaluation of the HLbL part of $a_\mu$ on the lattice has been achieved recently \cite{Blum:2014oka, Green:2015sra, Blum:2015gfa}, as well (see Ref.~\cite{Meyer:2018til} for a review on this topic).} As a result of this activity, the error on the $\pi^0$-, $\eta$-, and $\eta'$-exchange contributions is $\lesssim 4\cdot10^{-11}$ and $\lesssim2\cdot10^{-11}$ for two-pion contributions. The next ones in size, but with similar errors, are the axial-vector contributions, whose study and evaluation is the aim of this paper.
 
\subsection{Axial-vector contributions to the muon anomalous magnetic moment}
Although the Landau-Yang theorem \cite{Landau:1948kw,Yang:1950rg} forbids the annihilation of a spin-one particle into a pair of real photons, axial-vector exchange contributions to the HLbL piece of $a_\mu$ are still possible, since at least one photon is off-shell in both axial-$\gamma^*$-$\gamma^*$ vertices in such a contribution. Still, the Landau-Yang theorem imposes nontrivial requirements on the symmetry structure of the involved form factors, as we will see.

Early estimates of the corresponding contributions were carried out both in the extended Nambu-Jona-Lasino model by Bijnens, Pallante and Prades \cite{Bijnens:1995cc, Bijnens:1995xf,Bijnens:2001cq} and by Hayakawa, Kinoshita and Sanda using hidden local symmetry Lagrangians \cite{Hayakawa:1995ps, Hayakawa:1996ki}. The first group obtained $a_\mu^{\textrm{HLbL};A}=(2.5\pm1.0)\cdot10^{-11}$, which includes the ballpark value $1.7\cdot10^{-11}$, given by the second group.

Later on, Melnikov and Vainshtein \cite{Melnikov:2003xd} derived operator product expansion (OPE) constraints on the hadronic light-by-light (HLbL) tensor and built a model where these were saturated by dropping the momentum dependence of the
singly virtual transition form factors, which increases the axial contributions to $a_\mu$. As a result, their evaluation, $a_\mu^{\textrm{HLbL};A}=(22\pm5)\cdot10^{-11}$, is an order of magnitude larger than the previous estimates. 

Recently, there have been a couple of new estimates for \amuA{}: that by Pauk and Vanderhaeghen~\cite{Pauk:2014rta}, that accounted for the $f_1(1285)$ and $f_1(1420)$ contributions, obtaining $a_\mu^{\textrm{HLbL};A}=(6.4\pm2.0)\cdot10^{-11}$, and that by Jegerlehner~\cite{Jegerlehner:2017gek}, where also the $a_1(1260)$ was included as an intermediate state, obtaining $a_\mu^{\textrm{HLbL};A}=(7.6\pm2.7)\cdot10^{-11}$. With the $a_1$ contribution of the order of $1.9\cdot10^{-11}$ \cite{Jegerlehner:2017gek}, their agreement is remarkable considering the marked differences among both approaches. The large discrepancy with respect to the result in Ref.~\cite{Melnikov:2003xd} was ascribed both to the constant form factor used in \cite{Melnikov:2003xd} at the external vertex, and to an apparent violation of the Landau-Yang theorem (Bose symmetry), that they claimed required the form factor employed in Ref.~\cite{Melnikov:2003xd} to be anstisymmetric in order to have vanishing amplitude for real photons. Such a claim is however false as we demonstrate in the following section.\footnote{We are aware that M.~Hoferichter and collaborators have reached a similar conclusion~\cite{MartinINTgm2:2019}: that Ref.~\cite{Melnikov:2003xd} does not violate Landau-Yang theorem.}

On the experimental side, very little information is available (which again is partly due to the Landau-Yang theorem). Noticeably, the L3 Collaboration at LEP measured the diphoton coupling to the $f_1(1285)$ and $f_1(1420)$ states using their decays to $\pi^+\pi^-\eta$ \cite{Achard:2001uu} and $K_SK^\pm\pi^\mp$ \cite{Achard:2007hm} products. In this case, one could study the energy dependence of a linear combination of form factors relevant for  \amuA{}. We will take this information into account in our work.

Noteworthy, according to the most recent evaluations of \amuA{} in Refs. \cite{Pauk:2014rta, Jegerlehner:2017gek} the axial-vector contributions have a very similar uncertainty ($\sim 3\cdot10^{-11}$) as the sum of the tensor and higher-scalar meson contributions \cite{Pauk:2014rta,Danilkin:2016hnh, Knecht:2018sci}. 
One main motivation of our work is to confirm or disfavor this observation, especially due to the model-dependency of the estimations and some overlooked ambiguities related to the off-shellness, which could make further refined studies of \amuA{} needed.
An additional relevant outcome of this study is to show the sensitivity of \amuA{} to the asymptotic behavior of the form factors. This will be useful towards achieving a reliable model-dependent error estimation of  this contribution.

The outline of this paper is as follows: in \cref{sec:AxTFF} we define our notation and conventions together with the central results, pointing to the origin of the mentioned ambiguities, and commenting on the controversy about the Landau-Yang theorem raised in Refs.~\cite{Pauk:2014rta,Jegerlehner:2017gek}. Then, we compute the axial-vector exchange contribution to $a_\mu^{\textrm{HLbL}}$ in \cref{sec:Contributiontoamu}, with details relegated to \cref{sec:appHLBL}.
After that, we evaluate numerically \amuA{} for the lowest-lying axial multiplet in \cref{sec:numerics}, assessing as well the impact of higher order effects in resonance chiral theory (R$\chi$T). Finally, we outline our conclusions in \cref{sec:concl}. Several appendices complete our discussion:  \cref{app:schouten} collects several useful relations derived from Schouten identity;  \cref{app:conv} includes four other basis (and their relation with ours) for the axial transition form factors, briefly commenting about short-distance constraints; \cref{app:phenoRChT} summarizes the treatment of  $U(3)$ flavor breaking corrections in R$\chi$T and discusses the determination of the model parameters using short-distance QCD constraints and phenomenological information from LEP; \cref{app:HO} shows in detail the estimate of higher orders in R$\chi$T; finally, \cref{app:ope} summarizes the implications of the OPE for the axial transition form factors.

\section{The axial transition form factors} \label{sec:AxTFF}

\subsection{Definitions and main results}\label{sec:AxTFFdef}

Based on parity, charge conjugation and hermiticity, the axial transition 
matrix element with the electromagnetic currents $j^{\mu}\equiv \sum_i \mathcal{Q}_i\bar{q}_i \gamma^{\mu}q_i$ with $\mathcal{Q}_i$ the $i$th quark charge, defined as
\begin{equation}
    i\int d^4x e^{iq_1\cdot x} \bra{0} T\{ j^{\mu}(x) j^{\nu}(0)\} \ket{a^{\tau}} \equiv \mathcal{M}^{\mu\nu\tau}\varepsilon_{A\,\tau},
\end{equation}
where $\langle \gamma^{*}(q_1)\gamma^{*}(q_2) | A(p_A) \rangle \equiv (2\pi)^4\delta^{(4)}(q_1+q_2-p_A)ie^2\mathcal{M}^{\mu\nu\tau}\varepsilon_{A\,\tau}\varepsilon_{1\mu}^*\varepsilon_{2\nu}^*$, can be generally written as\footnote{$\mathcal{M}^{\mu\nu\tau}$ has GeV dimensions; $B_i$ and $C$, GeV$^{-2}$; $A$ is dimensionless. We use the notation $\epsilon_{\mu\nu\rho\sigma}p_i^{\rho}\equiv \epsilon_{\mu\nu p_i \sigma}$.}
\begin{equation}\label{eq:GenAxTFF}
   \mathcal{M}^{\mu\nu\tau}=
    i\epsilon^{\mu\nu\tau}{}_{\alpha}(q_1^{\alpha}A-q_2^{\alpha}\bar{A})
   +i\epsilon^{q_1q_2\alpha\tau}\left[ 
        g^{\mu}_{\alpha}(q_1^{\nu}B_1 +q_2^{\nu}B_2 ) -g^{\nu}_{\alpha}(q_1^{\mu}\bar{B}_2 +q_2^{\mu}\bar{B}_1 )
        \right]
   +i\epsilon^{\mu\nu q_1q_2} (q_1^{\tau}C +q_2^{\tau}\bar{C}),
\end{equation}
with $\epsilon^{0123}=+1$ and where, given a form factor $F\equiv F(q_1^2,q_2^2)$, we define $\bar{F}\equiv F(q_2^2,q_1^2)$. Note in addition that $(q_1 +q_2)\!\cdot\! \varepsilon_A =0$ implies that only the antisymmetric part in $C$ survives on-shell---yet we still keep it for later convenience. Defined in this way, only the tensor structure associated to $C$ ensures gauge invariance by itself. For the remaining set, gauge invariance implies\footnote{Lacking massless particles, all the form factors should be regular at $q_{1,2}^2=0$, implying that $2A(q_1^2,0) +(q_{12}^2 -q_1^2)B_1(q_1^2,0)=0$, while $B_2$ is not constrained for vanishing $q_{i}^2$.}
\begin{equation}\label{eq:gi}
    A +(q_1\!\cdot\!q_2)B_1 +q_2^2 B_2 = \bar{A} +(q_1\!\cdot\!q_2)\bar{B}_1 +q_1^2 \bar{B}_2 = 0. 
\end{equation}
However, these transition form factors (TFFs) are not independent of $C$---they are related via Schouten identitites (see \cref{app:schouten}). This implies that we can dismiss either $A, B_{1,2}$ and remove it from the equations above; in the following, we relegate $B_1$ and obtain $A=-q_2^2 B_2$.\footnote{In order to connect to alternative descriptions, where $B_1$ is not relegated, our choice is equivalent to shift $\Delta C= B_1$, $\Delta A = (q_1\cdot q_2)\overline{B}_1 +q_2^2 B_1$, $\Delta B_2 = -\overline{B}_1$ and analogously for the barred form factors.} Introducing $q_{12}=q_1 +q_2$, $\bar{q}_{12}=q_1-q_2$, and $C_{A(S)} = (C \mp\bar{C})/2$ ($B_{2A,2S}$ can be defined analogously), our expression for the TFFs parametrization can be expressed as
\begin{equation}\label{eq:axialFF}
    \mathcal{M}^{\mu\nu\tau}=
    i\epsilon^{\mu\alpha\tau q_1}(q_{2\,\alpha}q_2^{\nu} -g^\nu_\alpha q_2^2) B_2
    +i\epsilon^{\nu\alpha\tau q_2}(q_{1\,\alpha}q_1^{\mu} -g^\mu_\alpha q_1^2) \bar{B}_2 
    +i\epsilon^{\mu\nu q_1q_2} (\bar{q}_{12}^{\tau}C_A +q_{12}^{\tau}C_S).
\end{equation}
This definition is that appearing (up to overall factors and the spurious addition of $C_S$) in Ref.~\cite{Kuhn:1979bb,Rudenko:2017bel,Milstein:2019yvz} and roughly in \cite{Cahn:1986qg}, where $C\to0$ is taken. The relation to alternative existing bases is given in \cref{app:conv}. However, a relevant comment is in order here: since the connection among bases involves $C_S\neq0$ terms, different bases will
behave differently when reconstructing the axial-vector contribution to the HLbL Green's function, unless $q_{12}^{\tau}$-terms vanish in such procedure. More important, such procedure is not unique and only the residue at the axial pole is well defined. As an example, that different propagators $(g^{\mu\nu}-q_{12}^{\mu}q_{12}^{\nu}/X)(q^2-m_A^2)^{-1}$ with either $X=q_{12}^2$ or $X=m_A^2$ have been used in Refs.~\cite{Melnikov:2003xd,Jegerlehner:2015stw} and \cite{Pauk:2014rta}, respectively. The potential effect of such ambiguities and the off-shell prescription might be large, as we find. To circumvent this issue, one needs a well-defined off-shell prescription. In our case, we opt to use R$\chi$T that, despite some caveats~\cite{Bijnens:2003rc,Peris:2006ds,Masjuan:2007ay,Masjuan:2008fr}, has proven to successfully describe QCD at low energies at the precision that this work requires and allows, at the same time, to work directly at the level of Green's functions, that is the required input in computing \amu{}.

An additional comment concerns Landau-Yang theorem that, contrary to the claims in \cite{Pauk:2014rta,Jegerlehner:2015stw} was not violated in Ref.~\cite{Melnikov:2003xd}. To see this, note that \cref{eq:axialFF} can be re-expressed in the language of Ref.~\cite{Melnikov:2003xd} as
\begin{equation}
    \mathcal{M}^{\mu\nu\tau}\varepsilon^*_{1\mu}\varepsilon^*_{2\nu}\varepsilon_{A\tau} = i [ \{q_2F_2\tilde{F}_1\varepsilon_A\} B_2 +\{q_1F_1\tilde{F}_2\varepsilon_A\} \bar{B}_2  -\{F_2\tilde{F}_1\}(\bar{q}_{12}\cdot\varepsilon_A) C_A  ],
\end{equation}
where $F_i^{\mu\nu} =q_i^{\mu}\varepsilon_i^{\nu}-q_i^{\nu}\varepsilon_i^{\mu}$ and $\tilde{F}_i^{\mu\nu} = \epsilon^{\mu\nu\rho\sigma}F_{\rho\sigma}/2$. In this form, it is straightforward to show that the results in Ref.~\cite{Melnikov:2003xd} are equivalent to use only nonvanishing $B_{2S}$ (symmetric) form factors when computing \amuA{}. As such, the form factor employed in \cite{Melnikov:2003xd} cannot be antisymmetric, as it was argued in \cite{Pauk:2014rta,Jegerlehner:2015stw}. Further, as we detail below, the corresponding helicity amplitudes vanish for real photons, fulfilling Landau-Yang theorem. Actually, it is worth emphasizing that the OPE constraints derived in \cite{Melnikov:2003xd} apply to the $B_{2S}$ form factor alone (see \cref{app:ope}), and as such they cannot be used to set constraints to the $C_A$ form factor, as it was done in \cite{Jegerlehner:2015stw,Jegerlehner:2017gek}.

\subsection{Helicity amplitudes and cross section}

In the following, it will be useful to quote the nonvanishing on-shell helicity amplitudes in order to make contact with experimental results\footnote{We employ $q_{1(2)}=(E_{1(2)},0,0,\pm \boldsymbol{q})$, with $2m_A E_{1(2)} = m_A^2 \pm(q_1^2 -q_2^2)$, $\varepsilon^{\pm}_{A} =\varepsilon^{\pm}_{1} = \varepsilon^{\mp}_{2}= \mp(0,1,\pm i,0)/\sqrt{2}$, $\varepsilon^{0}_{1(2)}=(\pm \boldsymbol{q},0,0,E_{1(2)})/\sqrt{q_{1(2)}^2}$, and $\varepsilon_A^0 = (0,0,0,1)$ as in Refs.~\cite{Poppe:1986dq,Budnev:1974de,Gao:2010qx}.}
\begin{align}
    \mathcal{M}^{\pm\pm0} ={}& \mp\epsilon^{0123}\left[ 
    q_2^2\frac{m_A^2 +q_1^2 -q_2^2}{2m_A}B_2 - q_1^2\frac{m_A^2 +q_2^2 -q_1^2}{2m_A}\bar{B}_2
    +2\boldsymbol{q}^2 m_A C_A\right], \\
     \mathcal{M}^{\pm0\pm} ={}& \pm\epsilon^{0123}\left[
     \frac{q_1^2q_2^2}{\sqrt{q_2^2}}\bar{B}_2 - \frac{q_1\cdot q_2}{\sqrt{q_2^2}}q_2^2B_2 \right], \quad 
     \mathcal{M}^{0\mp\pm} ={} \pm\epsilon^{0123}\left[
     \frac{q_1^2q_2^2}{\sqrt{q_1^2}}B_2 - \frac{q_1\cdot q_2}{\sqrt{q_1^2}}q_1^2\bar{B}_2 \right],
\end{align}
where $\boldsymbol{q}= \lambda^{1/2}(m_A^2,q_1^2,q_2^2)(2m_A)^{-1} = [(q_1\cdot q_2) -q_1^2 q_2^2]^{1/2}m_A^{-1}$ refers to the photon momentum in the axial-vector meson rest frame---find similar results in Ref.~\cite{Milstein:2019yvz}. Note that the amplitudes vanish for $q_1^2=q_2^2=0$, in accordance to Landau-Yang theorem. A particularly interesting result is the cross section for $\gamma^*\gamma^*\to A$ that is relevant for $e^+e^-\to e^+e^- A$ production. Following the definitions in \cite{Budnev:1974de,Schuler:1997yw,Pascalutsa:2012pr,Milstein:2019yvz}, we find that 
\begin{align}
    \sigma_{TT} = \frac{1}{4 m_A \boldsymbol{q}}\pi\delta(s-m_A^2)|\mathcal{M}^{\pm\pm0}|^2, \quad
    \sigma_{TL} = \frac{1}{2 m_A \boldsymbol{q}}\pi\delta(s-m_A^2)|\mathcal{M}^{\pm\pm0}|^2, 
\end{align}
that, in the $q_1^2\to 0$ limit, produces a cross section $\sigma_{\gamma\gamma} \simeq \sigma_{LT} +\sigma_{TT}$ (find details in \cite{Schuler:1997yw,Pascalutsa:2012pr}) 
\begin{equation}\label{eq:sigmaGG}
    \sigma_{\gamma\gamma} = \delta(s -m_A^2)16\pi^2\frac{3\tilde{\Gamma}_{\gamma\gamma}}{m_A}x(1+x) \left[
    |\tilde{B}_2|^2 x\left(1 +\frac{x}{2}\right) +\frac{1}{2}|\tilde{C}_A|^2(1+x)^2 -x(1+x)\operatorname{Re}\tilde{B}_2\tilde{C}_A^* \right],
\end{equation}
where $x=Q_2^2/m_A^2$, $\tilde{B}_2(\tilde{C}_A) = B_2(C_A)/B_2(0,0)$, and where
\begin{equation}\label{eq:Awidth}
    \tilde{\Gamma}_{\gamma\gamma} = \lim_{Q^2_{1,2}\to 0} \frac{1}{2}\frac{m_A^2}{Q_2^2} \left[\Gamma_{TL} = \frac{1}{3}\frac{1}{2m_A} \int d{\Pi}_{\gamma\gamma} \sum_{T=\pm}|e^2\mathcal{M}^{T0T}|^2  \right]
                   = \frac{\pi \alpha^2}{12}m_A^5 |B_2(0,0)|^2.
\end{equation}
Note that in the narrow-width approximation $\pi \delta(s-m_A^2)\leftrightarrow m_a\Gamma_A[(s-m_A)^2 +m_A^2\Gamma_A^2]^{-1}$, that allows comparison to Ref.~\cite{Achard:2001uu}, (see Eqs.~(1-3) therein). The bracketed expression in \cref{eq:sigmaGG} compares to that for the simplified model (i.e., with $C_A=0$, see also \cref{app:QMbasis}) in Ref.~\cite{Achard:2001uu}, namely $ x(1+x/2)|F(Q^2)|^2$. For a dipole form factor $F(Q^2)$, Ref.~\cite{Achard:2001uu} finds a reasonable fit to data, suggesting that singly virtual form factors should not grow faster than $Q^{-4}$, that has relevant implications as we shall see.

\subsection{Form factors in \texorpdfstring{R\boldmath{$\chi$}T}{RChPT}}
Resonance chiral theory \cite{Ecker:1988te, Ecker:1989yg} was developed  with the purpose of---on the one hand---enlarging the domain of applicability of Chiral Perturbation Theory ($\chi$PT) \cite{Weinberg:1978kz, Gasser:1983yg, Gasser:1984gg} to higher energies and---on the other---to explain the values of the subleading  low-energy  $\chi$PT constants in terms of the lightest  meson masses and couplings; based only upon the approximate unitary flavor symmetry for the resonances,  and recovering $\chi$PT at low enough energies. It has also been systematically applied to long-distance dominated kaon decays (see \cite{Cirigliano:2011ny} for a review), hadronic tau meson decays (remarkably, the Monte Carlo based on R$\chi$T results~\cite{Dumm:2009va, Shekhovtsova:2012ra, Nugent:2013hxa} provides the best performance in describing three-prong pion $\tau$ decays in $R_{D^*}$ analyses~\cite{Aaij:2017uff, Aaij:2017deq}), and to the study of Green functions which are order parameters of chiral symmetry breaking (see, for instance, \cite{Cirigliano:2006hb, Kampf:2011ty}). Unitarized R$\chi$T has also been applied succesfully in the timelike region to understand meson-meson scattering  (see e.g. Refs. \cite{Jamin:2000wn, Guo:2011pa}). In the spacelike region, it has been employed satisfactorily to explain the $\pi^0$ \cite{Kampf:2011ty}, $\eta$ and $\eta'$ transition form factors \cite{Roig:2014uja} and their corresponding contributions to $a_\mu^{\textrm{HLbL};P}$. Variants and corrections to this approach have been considered in Refs. \cite{Czyz:2012nq, Guevara:2018rhj}. Figures 4 and 5 in the latter reference show a good fit to data on the $\pi^0$, $\eta$ and $\eta'$ transition form factors with $a_\mu^{\textrm{HLbL};P}$ in the ballpark of the results in the  literature. As such, we expect that a similar performance is obtained for the axial case.
 
 Using the R$\chi$T Lagrangian \cite{Ecker:1988te, Ecker:1989yg} that saturates the $\mathcal{O}(p^6)$ LECs in the odd-intrinsic parity sector \cite{Kampf:2011ty}, the leading contribution can be conveniently expressed as
\begin{align}
    \mathcal{M}^{\mu\nu\tau}\varepsilon_{A\tau} ={}& \sum_V c_{AV} 
    \frac{(q_1^2 -q_2^2) \bra{0} A_{\rho\lambda} \ket{A}}{(q_1^2 -M_V^2)(q_2^2 -M_V^2)}
    \left([\epsilon^{\mu\nu\rho q_2}q_1^\lambda + \epsilon^{\mu\nu\rho q_1}q_2^\lambda] -[\epsilon^{\nu\rho q_1 q_2}g^{\mu\lambda} +\epsilon^{\mu\rho q_1 q_2}g^{\nu\lambda}]
        \right) \nonumber \\
    \equiv{}& \mathcal{M}^{\mu\nu;\rho\lambda}_A \bra{0} A_{\rho\lambda} \ket{A}, 
    \qquad\qquad\qquad [c_{AV}= -2\sqrt{2}\kappa_5^{AV} F_V \operatorname{tr}(\{ V,A\}Q)\operatorname{tr}(VQ)]. \label{eq:RChThaxialtoGG}
\end{align}
The equation above will be the central quantity in determining the axial-vector contribution to the HLbL Green's function (and thereby $a_{\mu}^{\textrm{HLbL};A}$) in the following section. In order to make contact with the axial TFFs, one substitutes the $\bra{0} A_{\rho\lambda} \ket{A}$ matrix element in the equation above, obtaining the following amplitude for the $A\to\gamma^*\gamma^*$ transition assuming ideal mixing and the isospin limit ($M_{\rho}=M_{\omega}$ $F_{\rho}=F_{\omega}$):
\begin{align}
    \mathcal{M}^{\mu\nu\tau} ={}&  \frac{2e^2 c_A M_{A}^{-1}(q_1^2 -q_2^2)}{ (q_1^2 -M_V^2)(q_2^2 -M_V^2)} \big(
    i\epsilon^{\mu\alpha\tau q_1}[q_2^{\nu}q_{2\alpha}-g^{\nu}_{\alpha}q_2^2] 
    - i\epsilon^{\nu\alpha\tau q_2}[q_1^{\mu}q_{1\alpha}-g^{\mu}_{\alpha}q_1^2]
    +i\epsilon^{\mu\nu q_1 q_2}\bar{q}_{12}^{\tau}
    \big),
\end{align}
where $V\to\rho\omega$ for $a_1$ and $f_1$ cases, while $V\to\phi$ for $f_1'$. Finally, the form factors read 
\begin{align}\label{eq:cAdef}
    C_A = B_2 = -\bar{B}_2 = \frac{2c_A}{M_A}\frac{q_1^2 -q_2^2}{(q_1^2 -M_V^2)(q_2^2 -M_V^2)}, \ \ 
    c_{(a_1,f_1,f_1^\prime)} = -\left( 1, \frac{5}{3}, \frac{\sqrt{2}}{3} \right)\frac{4F_{V}\kappa_5^{VA}}{3}.
\end{align}
Note the absence of symmetric form factors, that are chirally suppressed and appear at higher orders (see \cref{sec:numerics} and \cref{app:HO}).
Although both, $F_{\rho\omega}$ and $F_\phi$ depart from $F_V$ by $\mathcal{O}(m_{\pi,K}^2)$ corrections \cite{Guevara:2018rhj}, when the appropriate short-distance constraints are required, one recovers $F_{\rho\omega}=F_\phi=F_V$ \cite{Guevara:2018rhj}. 
We note that the ideal mixing for the spin-one nonets that we have used (that is predicted with $N_C\to\infty$) is supported by the fact that $\textrm{BR}(f_1\to\phi \gamma)=(7.4\pm2.6)\cdot10^{-4} \ll \textrm{BR}(f_1\to\rho
\gamma)=(5.3\pm1.2)\cdot10^{-2}$. For the numerical inputs, we refer to \cref{app:phenoRChT}.

\subsection{Additional vector multiplets in \texorpdfstring{R\boldmath{$\chi$}T}{RChPT}}

As a result of their antisymmetric nature, the singly virtual form factors in \cref{eq:cAdef} will behave as a constant for large spacelike virtualities, while the results from L3 Collaboration suggest a $Q^{-4}$ behavior (see \cref{sec:AxTFFdef}). Such behavior requires the inclusion of additional resonances. With an additional multiplet satisfying the condition $F_{V'} \kappa_5^{V'A}=-F_V \kappa_5^{VA}(M_{V'}^2/M_V^2)$, the asymptotic behavior is improved but it is not yet satisfactory. However, with a third multiplet fulfilling
\begin{equation} \label{relations3V}
 F_V \kappa_5^{VA}+F_{V'} \kappa_5^{V'A}+F_{V''} \kappa_5^{V''A}=0\,,\quad F_{V'} \kappa_5^{V'A}=\frac{F_V  \kappa_5^{VA}(M_V^2-M_{V''}^2)M_{V'}^2}{ (M_{V''}^2-M_{V'}^2)M_V^2}\,,
\end{equation}
the asymptotic behavior can be considered realistic.\footnote{This observation is suggested by the two data points measured in the region $Q^2\in[0.6,4]$ GeV$^2$ by the L3 Collaboration \cite{Achard:2001uu} It is hard to draw any conclusion on this issue from Ref. \cite{Achard:2007hm}, as both $\eta(1475)$ and $f_1(1420)$ states are required to describe the data.}. The previous equation \footnote{We note that these relations will change beyond tree-level, when renormalized couplings are considered. This issue has been studied e.g. in Ref.\cite{Rosell:2004mn} for the pion vector form factor.} fixes the relevant combinations $F_{V'} \kappa_5^{V'A}$ and $F_{V''} \kappa_5^{V''A}$ in terms of $F_V \kappa_5^{VA}$ and the masses of the vector multiplets, which are known phenomenologically, as it is discussed in \cref{app:phenoRChT}. We note that we have to ensure the normalization of these form factors (with one, two, or three vector resonance multiplets) at zero photon virtualities be the same. This is achieved if the form factor with two vector nonets is multiplied by the factor $M_{V'}^2/M_V^2/(-1+M_{V'}^2/M_V^2)$ and the one with three vector multiplets by $M_{V'}^2 M_{V''}^2/(M_{V'}^2-M_V^2)/(M_{V''}^2-M_V^2)$.

\section{Axial contribution to \texorpdfstring{\boldmath{$a_{\mu}^{\textrm{HLbL}}$}}{the (g-2) HLbL} in \texorpdfstring{R\boldmath{$\chi$}T}{RChPT}} \label{sec:Contributiontoamu}

\begin{figure}
    \centering
    \includegraphics[width=\textwidth]{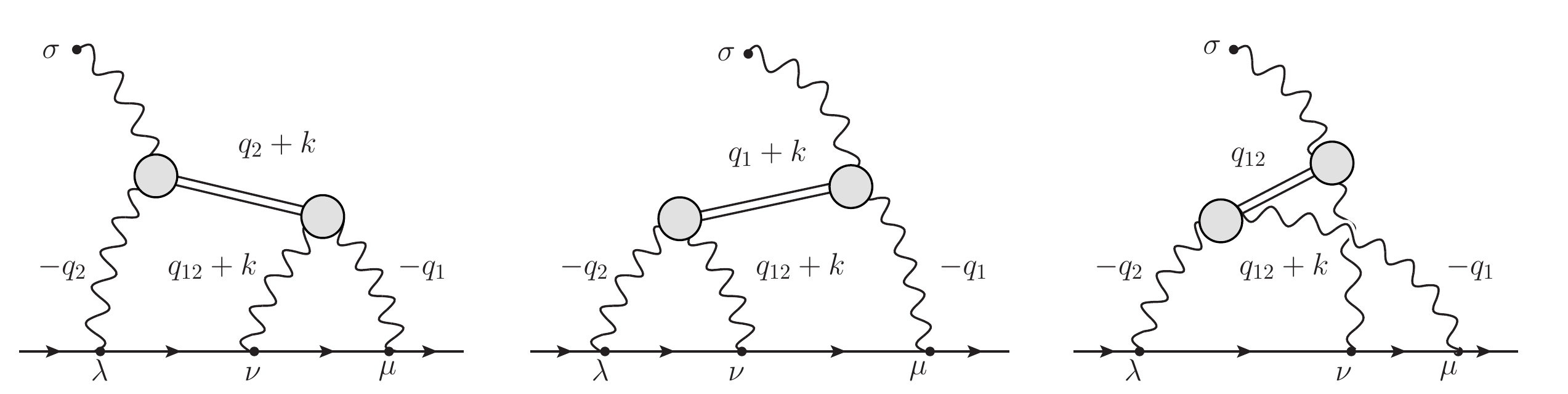}
    \caption{The $s$-, $t$-, and $u$-channel resonance exchange (depicted with a double line) contribution to the HLbL. The momentum flows out from the blobs except for the external incoming momentum $k$.}
    \label{fig:hlblnew}
\end{figure}

The HLbL contribution to $a_{\mu}$  in the vanishing external momentum limit can be obtained using the  projection techniques outlined in Refs.~\cite{Knecht:2001qf,Knecht:2018sci}.\footnote{We note that here we already perform the change of variables $q_1\to-q_1$, $q_2\to q_{12} +k$ at the matrix element level as compared to Refs.~\cite{Knecht:2001qf,Knecht:2018sci}, where this is performed in a second stage.} Particularly, one finds~\cite{Knecht:2001qf,Knecht:2018sci}
\begin{equation}
    a_{\mu} = \frac{1}{48m_{\mu}}\operatorname{tr}(\slashed{p}+m_{\mu})[\gamma^{\rho},\gamma^{\sigma}](\slashed{p}+m_{\mu})\Gamma_{\rho\sigma}(p,p),
\end{equation}
with 
\begin{align}
    \Gamma^{\rho\sigma}(p,p) = -ie^6\! \int \frac{d^4q_1}{(2\pi)^4}\frac{d^4q_2}{(2\pi)^4} \frac{\gamma_{\mu} (\slashed{p}+\slashed{q}_1 \!+ \!m_{\mu}) \gamma_{\nu} (\slashed{p}-\slashed{q}_2 \!+\!m)\gamma_{\lambda}}{q_1^2 q_2^2 q_{12}^2[(p+q_1)^2 -m_{\mu}^2][(p-q_2)^2 -m_{\mu}^2]} \partial\Pi_{\textrm{HLbL}}^{\mu\nu\lambda\sigma\rho}(q_1,q_2).
\end{align}
In the previous equation we have introduced $\partial\Pi_{\textrm{HLbL}}^{\mu\nu\lambda\sigma\rho}(q_1,q_2) \equiv \lim_{k\to0} (\partial/\partial_{k_{\rho}})\Pi_{\textrm{HLbL}}^{\mu\nu\lambda\sigma}(-q_1,q_{12}+k,-q_2,-k)$, where $\Pi_{\textrm{HLbL}}^{\mu\nu\lambda\sigma}$ stands for the HLbL tensor. In the case of axial-mesons (see \cref{fig:hlblnew}), and after dropping irrelevant $k$ terms, this reads
\begin{align}
    \Pi_{\textrm{HLbL};A}^{\mu\nu\lambda\sigma}(-q_1,q_{12},-q_2,-k) ={}& 
    i\mathcal{M}_A^{\mu\nu;\alpha\beta}(-q_1,q_{12}) i\Delta_F^R(q_{2})_{\alpha\beta,\bar{\alpha}\bar{\beta}} i\mathcal{M}_A^{\lambda\sigma;\bar{\alpha}\bar{\beta}}(-q_{2},-k)   \nonumber\\ {}&
    +i\mathcal{M}_A^{\lambda\nu;\alpha\beta}(-q_{2},q_{12}) i\Delta_F^R(q_1)_{\alpha\beta,\bar{\alpha}\bar{\beta}} i\mathcal{M}_A^{\mu\sigma;\bar{\alpha}\bar{\beta}}(-q_1,-k) \nonumber\\ {}&
    +i\mathcal{M}_A^{\mu\lambda;\alpha\beta}(-q_1,-q_2) i\Delta_F^R(q_{12})_{\alpha\beta,\bar{\alpha}\bar{\beta}} i\mathcal{M}_A^{\nu\sigma;\bar{\alpha}\bar{\beta}}(q_{12},-k),        
\end{align}
with $\mathcal{M}_A^{\mu\nu;\rho\sigma}$ defined in \cref{eq:RChThaxialtoGG} and $\Delta_F^R(q)_{\alpha\beta,\bar{\alpha}\bar{\beta}}$ standing for the resonance propagator,\footnote{As we advanced, one advantage of the Lagrangian formalism is that there is no ambiguity when computing Green's functions. In our case, we choose to represent the spin-one resonances by antisymmetric tensor fields, so the corresponding propagators can be read from \cref{Spin1propAS}. While physical  observables are independent of our choice for representing the (axial)-vector meson fields, this does not need to be the case for individual contributions to them if asymptotic constraints are not properly taken into account, and deserves further study in the context of $a_\mu$ (see e. g. Refs. \cite{Ecker:1989yg, GomezDumm:2000fz}).}
\begin{equation}\label{Spin1propAS}
    \Delta_F^R(q)^{\mu\nu,\rho\sigma} = -\frac{[g^{\mu\rho}q^{\nu}q^{\sigma} -g^{\mu\sigma}q^{\nu}q^{\rho}  +g^{\mu\rho} g^{\nu\sigma}(M_R^2 -q^2)] -(\mu\leftrightarrow\nu)}{(q^2 -M_R^2)M_R^2},
\end{equation}
leading to 
\begin{align}\label{eq:aHLbLtensor}
\partial\Pi_{\textrm{HLbL};A}^{\mu\nu\lambda\sigma\rho} = 
    i\Delta_F^R(q_2)_{\alpha\beta;\bar{\alpha}\bar{\beta}}F_A(q_1^2,q_{12}^2)F_A(q_2^2,0) \big[
      (\epsilon^{\lambda\sigma\bar{\alpha}\rho}q_{2}^{\bar{\beta}} +\epsilon^{\lambda\sigma\bar{\alpha}q_2}g^{\rho\bar{\beta}}) +
      (\epsilon^{\sigma\bar{\alpha}\rho q_2}g^{\lambda\bar{\beta}} 
      +\epsilon^{\lambda\bar{\alpha}\rho q_2}g^{\sigma\bar{\beta}})
    \big] \nonumber\\ \times
    \big[
      (\epsilon^{\mu\nu\alpha q_{12}}q_{1}^{\beta} +\epsilon^{\mu\nu\alpha q_{1}}q_{12}^{\beta}) -
      (\epsilon^{\nu\alpha q_1 q_2}g^{\mu\beta} +\epsilon^{\mu\alpha q_1 q_2}g^{\nu\beta} )
    \big] \qquad + (^{\mu\leftrightarrow\lambda}_{q_1\leftrightarrow q_2}) \nonumber\\ \qquad
+   i\Delta_F^R(q_{12})_{\alpha\beta;\bar{\alpha}\bar{\beta}}F_A(q_1^2,q_{2}^2)F_A(q_{12}^2,0) \big[
      (\epsilon^{\nu\sigma\bar{\alpha}\rho}q_{12}^{\bar{\beta}} +\epsilon^{\nu\sigma\bar{\alpha}q_{12}}g^{\rho\bar{\beta}}) +
      (\epsilon^{\sigma\bar{\alpha}\rho q_{12}}g^{\nu\bar{\beta}} 
      +\epsilon^{\nu\bar{\alpha}\rho q_{12}}g^{\sigma\bar{\beta}})
    \big] \nonumber\\ \times
    \big[
      (\epsilon^{\mu\lambda\alpha q_{2}}q_{1}^{\beta} +\epsilon^{\mu\lambda\alpha q_{1}}q_{2}^{\beta}) -
      (\epsilon^{\lambda\alpha q_1 q_2}g^{\mu\beta} +\epsilon^{\mu\alpha q_1 q_2}g^{\lambda\beta} )
    \big],    
\end{align}
where $F_A(q_1^2,q_2^2) = c_A(q_1^2 -q_2^2)(q_1^2 -M_V^2)^{-1}(q_2^2 -M_V^2)^{-1}$, see \cref{eq:cAdef}. Following the method of Gegenbauer polynomials in Ref.~\cite{Jegerlehner:2009ry} to evaluate the integral, one can show that \begin{align}
    a_{\mu} ={}& \left(\frac{\alpha}{\pi}\right)^3 \frac{2\pi}{3} \int dt  dQ_1 dQ_2 \left[
    \sqrt{1-t^2} \frac{Q_1^3 Q_2^3}{Q_{12}^2 m_{\mu}^2} \sum_{i=1}^2 K_i(Q_1^2,Q_2^2,t) ,
    \equiv \sum_{i=1}^2 w_i\right]\label{eq:lblmaster}
\end{align}
where the expressions for $K_i(Q_1^2,Q_2^2,t)$ are given in \cref{sec:appHLBL}.

\section{Numerical evaluation of \amuA{}} \label{sec:numerics}

We evaluate \amuA{} including the contribution of the lightest axial-vector multiplet with up to three vector multiplets for the reasons discussed below. The results are obtained upon numerical integration  of the formulas derived in \cref{sec:Contributiontoamu}. One very important thing to note is that the first contribution, given by the integration of \cref{eq:lblmaster}, is not convergent for only one vector multiplet (find comments on this aspect in \cref{sec:appHLBL}). Because of this, we will only quote our results for either two ($2Vs$) or three ($3Vs$) vector multiplets. We do not consider excited axial-vector multiplets as there is only one unambiguous nonet in the PDG \cite{Tanabashi:2018oca}.\footnote{One could, in principle, explore the impact of the infinite tower of states predicted in the large-$N_C$ limit by means of Regge models, see e. g. Ref. \cite{Colangelo:2019uex}.} Although we consider the latter our preferred result, as its asymptotic behavior seems to agree with the trend shown by L3 data (see \cref{sec:AxTFFdef}), it is nevertheless informative to compare both values and to verify that a more realistic (stronger) asymptotic damping of the relevant form factors yields smaller contributions with three vector multiplets than with only two. Moreover, as we discuss in the following, we will use the resulting difference as an error estimate.\\
In our evaluation, we are using the restrictions in \cref{relations3V} (and their analogous for only two vector multiplets) that link the couplings and masses of the different multiplets. As a result of this, to obtain the errors we will float $F_V\kappa_5^{VA}$, $M_V$ and $M_A$ independently, but assume $M_{V^\prime}$ and $M_{V^{\prime\prime}}$ to be fully correlated with $M_V$. The final error is the combination in quadrature of them. Our results are summarized in \cref{Table}, where the different axial-vector meson contributions, in units of $10^{-11}$, are given.
\begin{table}[h]
\begin{center} 
\begin{tabular}{ |c|c|c|c| } 
 \hline
 Vector multiplets & $a_1$ & $f_1$ & $f_1^\prime$\\
 \hline
 Two & $1.13^{+0.21}_{-0.22}$ & $3.14^{+0.58}_{-0.61}$ & $0.07\pm0.04$ \\
 \hline
 Three & $0.21\pm0.04$ & $0.58\pm0.11$ & $0.015\pm0.008$ \\ 
 \hline
\end{tabular}
\caption{Different axial-vector meson contributions to \amu{} in in units of $10^{-11}$. The labels for the second and third row stand for the number of vector multiplets entering the form factor description.}
\label{Table}
\end{center}
\end{table}
Adding up the individual contributions, we obtain
\begin{equation} \label{finalresults}
a_\mu^{a_1+f_1+f_1^\prime,2Vs}=(4.34^{+0.62}_{-0.65})\cdot10^{-11}\,,\quad     a_\mu^{a_1+f_1+f_1^\prime,3Vs}=(0.81\pm0.12)\cdot10^{-11}\,.
\end{equation}
We observe that, while our result with two vector multiplets lies in between the early~\cite{Bijnens:1995cc, Bijnens:1995xf,Bijnens:2001cq, Hayakawa:1996ki,Hayakawa:1997rq} and most recent~\cite{Jegerlehner:2017gek, Pauk:2014rta} evaluations, it reveals a much smaller value than all preceding analysis (yet in line with early studies) when three vector multiplets are included. We emphasize that such choice has been adopted in order to satisfy the leading power of the asymptotic behavior suggested by the last two L3 data points  \cite{Achard:2001uu}, and as such represents our preferred value. Still, this points out the need of additional data at high energies and a more refined analysis regarding the form factor description there. 

Finally, one might wonder about the effect of higher order corrections in R$\chi$T. Especially regarding the appearance of a symmetric form factor, that was conjectured to play the main role earlier~\cite{Pauk:2014rta}, and whether the differences we found could be ascribed to it. To that object, we estimate such contribution in \cref{app:HO}, finding that, despite our result is analogous to that in Ref.~\cite{Pauk:2014rta} when on-shell, it turns out to be much smaller than \cite{Pauk:2014rta}, of the order of \cref{finalresults} and with opposite sign (such hierarchy was expected due to the chiral suppression). This illustrates again the problems that may arise when naively reconstructing Green's functions using naive propagators together with on-shell form factors. Incorporating these corrections as an additional uncertainty, we obtain 
\begin{equation}
   a_\mu^{a_1+f_1+f_1^\prime}=(0.8^{+3.5}_{-0.8})\cdot10^{-11}\, ,
\end{equation}
that is also in line with a recent result in Ref.~\cite{Dorokhov:2019tjc}. As a final comment, new publications have found far larger results for axial contributions~\cite{Leutgeb:2019gbz,Cappiello:2019hwh}. We emphasize that this is related to the way certain short-distance constraints~\cite{Melnikov:2003xd} for the longitudinal contribution are fulfilled, that is still work in progress and is missing insofar in our approach, see \cref{app:VVA,app:HO}. If confirmed by future (dispersive, lattice, etc.) studies, our finding would imply that axial contributions turn out to be similar in size to the sum of tensor and higher-scalar contributions, with an error that is negligible at the current level of requested accuracy, that underlines the need for further studies regarding the axial contributions to \amu{}.

\section{Conclusions}\label{sec:concl}

In this article, we have studied the axial-vector contributions to the hadronic light-by-light piece of the muon anomalous magnetic moment, \amuA{}. This is a timely enterprise, as we are eagerly awaiting the first publication from the Muon g-2 FNAL Collaboration, which would give $a_\mu$ with a comparable uncertainty to the LBNL measurement. In the years to come, FNAL will reach a fourfold improved uncertainty which will challenge our understanding of the Standard Model and its possible extensions provided a similar reduction can be achieved on the theory prediction, that is dominated by hadronic uncertainties. In fact, the spectacular improvement on the accuracy of the HVP evaluations demands a deeper understanding of the hadronic light-by-light piece, wherein the lightest pole cuts are already known with enough precision. Therefore, subleading contributions which are---however---subject to comparatively large uncertainties, become relevant for this endeavor, and the large relative error of these (otherwise small) contributions coming from heavier intermediate states in the HLbL diagrams needs to be reduced. In this context, we have studied the axial-vector contributions to \amuA{} within R$\chi$T. Our most important results  are discussed in the following.

We have motivated our conventions for the relevant matrix element and related ours with others  employed before in the literature, clarifying existing controversies, identifying previous theoretical ambiguities, and providing a dictionary to translate from one basis to another. As there are not many studies of this particular topic and a unified treatment has not been adopted yet, we believe our paper can constitute a reference in this respect. Further, this is, to our best knowledge, the first derivation for \amuA{} within R$\chi$T, and might be an useful reference for future studies.

As opposed to previous approaches, we have employed a Lagrangian formalism, that allows to work directly at the level of Green's functions, that is the required ingredient in evaluating \amuA{}. At this respect, the first important finding is the large model dependency that naively reconstructed Green's functions employed in previous studies might have, since only their residue at the axial pole is model-independent. This has been neatly illustrated in the context of R$\chi$T. The second important finding is the impact of the asymptotic behavior demanded to the relevant form factors. In particular, the comparison of our two evaluations in \cref{finalresults} shows neatly that the main systematic  uncertainty comes from the lack of data probing the asymptotic region of the axial transition form factors. Therefore, it would be crucial that a number of data points at large $Q^2$ were measured for the $e^+e^-\to e^+e^- A$ cross-section. An interesting and complementary study would be to address the $e^+e^-\to f_1$ production, that has been recently measured by SND Collaboration~\cite{Achasov:2019wtd}. Finally, it might also be interesting to study the sum rules as discussed in \cite{MartinINTgm2:2019}.

In addition to this, dispersive and lattice evaluations of \amuA{} would contribute to the understanding of these contributions and to reduce the corresponding  uncertainty in the SM prediction of $a_\mu$.

{\textbf{\textit{Note added:}}} After the publication of this work, Ref.~\cite{Hoferichter:2020lap} appeared, that shows that our $C_A(q_1^2,q_2^2)$ form factor behaves, asymptotically, as $q_i^{-6}$, that supports choosing the three vector multiplets results in our evaluation. For comparison, we note that their $F_i^A$ form factors identify with ours as $F_1=m_A^2 C_A$, $F_2= m_A^2 \bar{B}_2$, $F_3= -m_A^2 B_2$.

\section*{Acknowledgments}
P.~S. acknowledges  K.~Kampf and P.~Masjuan for discussions. 
The work of P.~S. is supported by Ministerio de Industria, Econom{\'ia} y Competitividad under the grant SEV-2016-0588, the grant 754510 (EU, H2020-MSCA-COFUND2016), and the grant FPA2017-86989-P, as well as by Secretaria d'Universitats i Recerca del
Departament d’Economia i Coneixement de la Generalitat de Catalunya under the grant
2017 SGR 1069. 
The work of P.~R.  has been partially funded by Conacyt through the project 250628 (Ciencia B\'asica).  The funding of Fondo SEP-Cinvestav 2018 (project number 142) is also acknowledged.

\appendix

\section{Schouten identities}\label{app:schouten}

It will be useful to employ the Schouten identity
\begin{equation}
      \epsilon^{\mu\nu\rho\sigma}g^{\lambda\delta} = 
        \epsilon^{\delta\nu\rho\sigma}g^{\lambda\mu} +\epsilon^{\mu\delta\rho\sigma}g^{\lambda\nu}
        +\epsilon^{\mu\nu\delta\sigma}g^{\lambda\rho} +\epsilon^{\mu\nu\rho\delta}g^{\lambda\sigma}
\end{equation}
in the following; particularly, the latter implies in our case that
\begin{equation}
      \epsilon^{\mu\nu q_1q_2}q_1^{\tau} = 
        \epsilon^{\tau\nu q_1q_2}q_1^{\mu} +\epsilon^{\mu\tau q_1q_2}q_1^{\nu}
        +\epsilon^{\mu\nu\tau q_2}q_1^{2} +\epsilon^{\mu\nu q_1\tau}(q_1 \cdot q_2), 
\end{equation}
and analogously for the $\mu\leftrightarrow\nu$, $1\leftrightarrow2$ expression. The former can be conveniently rewritten in terms of gauge invariant terms for later convenience (meaning they are orthogonal to $q_1^{\mu}$ and $q_2^{\nu}$) in different ways:
\begin{align}
      \epsilon^{\mu\nu q_1q_2}q_1^{\tau} ={}&
        \epsilon^{\nu\alpha\tau q_2}(q_1^{\alpha}q_1^{\mu} -g^{\mu\alpha}q_1^2)
       +\frac{q_1\cdot q_2}{q_2^2}\epsilon^{\mu\alpha\tau q_1}(q_2^{\alpha}q_2^{\nu} -g^{\nu\alpha}q_2^2)
       +\frac{\epsilon^{\mu q_2\tau q_1}}{q_2^2}[q_1^{\nu}q_2^2 -q_2^{\nu}(q_1\!\cdot\! q_2)], \\
      \epsilon^{\mu\nu q_1q_2}q_1^{\tau} ={}&
        \epsilon^{\nu\alpha\tau q_2}(q_1^{\alpha}q_1^{\mu} -g^{\mu\alpha}q_1^2)
       +\epsilon^{\mu \alpha \tau q_1}[q_1^{\nu}q_2^{\alpha} -g^{\nu\alpha}(q_1\cdot q_2)],\\
      \epsilon^{\mu\nu q_1q_2}q_1^{\tau} ={}&
        \epsilon^{\nu\alpha\tau q_2}(q_1^{\alpha}q_1^{\mu} -g^{\mu\alpha}q_1^2)
       -\epsilon^{\mu \alpha \tau q_1}(q_2^{\nu}q_2^{\alpha} -g^{\nu\alpha}q_2^2)
       +\epsilon^{\mu\alpha\tau q_1}[q_2^{\alpha}q_{12}^{\nu} -g^{\nu\alpha}(q_2\cdot q_{12})], \label{eq:schoutenRCHT}
\end{align}
and, again, the corresponding $\mu\leftrightarrow\nu$, $1\leftrightarrow2$ expressions. All of them allow to relate the different possible parametrizations of the axial TFFs. An additional interesting result, that has been used in \cite{Kampf:2011ty}, is the following
\begin{equation}
   \epsilon_{\mu\nu\alpha\beta}g_{\rho\sigma}\langle \{V^{\mu\nu},A^{\alpha\rho} \} f_+^{\beta\sigma}\rangle
   =\epsilon_{\mu\nu\alpha\beta}g_{\rho\sigma}\langle \{f_+^{\alpha\rho}, V^{\beta\sigma} \} A^{\mu\nu} \rangle. 
\end{equation}
Note this implies that, exchanging $V\leftrightarrow f_+$ leads to the same term up to a sign. Further, for $V\propto f_+$, it vanishes, having no contribution to external vector currents nor the presence of a two-resonance term.

\section{Other bases for the axial transition form factor}\label{app:conv}

\subsection{Helicity basis I}\label{app:JegerBasis}

A popular choice adopted in Refs.~\cite{Kopp:1973hp,Greco:1977pu,Renard:1983xn},\footnote{Such a choice is equivalent to use the Schouten identities to get rid of $A$; then, the Ward identities imply $(q_1\cdot q_2)B_1 +q_2^2B_2=0 \to \{ B_1= -q_2^2B, B_2=(q_1\cdot q_2)B \}$, that carries the $q_{i}^2$ suppression we find in \cref{eq:FredTrans}.  In order not to have it, one would require $B_1 = -q_2^2 (q_1\!\cdot\! q_2)^{-1} B_2$. Note that such additional suppression artificially implies that only the antisymmetric $F_A$ term contributes to $a_{\mu}^{\textrm{HLbL};A}$~\cite{Jegerlehner:2017gek}. This is clearly artificial and shows that such basis is not an optimal choice, for it introduces artificial kinematical zeroes. Further, the form factor in \cite{Jegerlehner:2017gek} is neither analytic nor antisymmetric at $q_1^2=q_2^2=0$.}  and used in Ref.~\cite{Jegerlehner:2015stw} to compute $(g-2)_{\mu}^{\textrm{HLbL};A}$, is
\begin{equation}\label{eq:tffJeger}
    \mathcal{M}^{\mu\nu\tau}=
    i\epsilon^{\nu q_1 \tau q_2}[q_2^{\mu}q_1^2 -q_1^{\mu}(q_1\!\cdot\!q_2) ]F_A'
    +i\epsilon^{\mu q_2 \tau q_1}[q_1^{\nu}q_2^2 -q_2^{\nu}(q_1\!\cdot\!q_2) ]F_A''
    +i\epsilon^{\mu\nu q_1q_2} (q_1 -q_2)^{\tau}\frac{1}{2}F_A.
\end{equation}
From the Schouten identities one can show the relations
\begin{align}
    C = \frac{F_A}{2} +q_2^2 F_A'';  \bar{C} = -\frac{F_A}{2} +q_1^2 F_A';  B_2 = -[q_1^2 F_A' +(q_1\!\cdot\!q_2)F_A''];  \bar{B}_2 = -[q_2^2 F_A'' +(q_1\!\cdot\!q_2)F_A']. \label{eq:FredTrans}
\end{align}

\subsection{Quark-model inspired}\label{app:QMbasis}

Another common choice is to take a single form factor~\cite{Cahn:1986qg,Cahn:1987zp,Achard:2001uu,Achard:2007hm,Pascalutsa:2012pr}:
\begin{equation}
    \mathcal{M}^{\mu\nu\tau} = i\epsilon^{\mu\nu\tau\alpha}(q_1^2q_{2\alpha} -q_2^2q_{1\alpha})A(q_1^2,q_2^2) 
    = i[\epsilon^{\nu\alpha\tau q_2}g_{\alpha}^{\mu}(-q_1^2) +\epsilon^{\mu\alpha\tau q_1}g^{\nu}_{\alpha}(-q_2^2)]  A(q_1^2,q_2^2).
\end{equation}
Note however that the formula above is not gauge invariant. The latter can be achieved via\footnote{While added terms are irrelevant when connecting to on-shell currents, such as in $e^+e^-$ production, this is not the case in $(g-2)$ where, in a general $R_{\xi}$ gauge, the photon propagator 
demands to keep those terms in order to obtain a $\xi$-independent result.}
\begin{equation}
    \mathcal{M}^{\mu\nu\tau} =
    i\epsilon^{\mu\alpha\tau q_1}(q_{2\,\alpha}q_2^\nu -g^\nu_\alpha q_2^2) A(q_1^2,q_2^2)
    +i\epsilon^{\nu\alpha\tau q_2}(q_{1\,\alpha}q_1^\mu -g^\mu_\alpha q_1^2) A(q_1^2,q_2^2),
\end{equation}
allowing to identify $B_2 =\bar{B}_2 = A(q_1^2,q_2^2)$. Particularly, it is the last one that was used in Ref.~\cite{Pauk:2014rta} to compute the contribution to $(g-2)$.

\subsection{Helicity basis II}

Finally, we find a different choice in Ref.~\cite{Poppe:1986dq,Pascalutsa:2012pr} based on helicities. Defining $X=(q_1\cdot q_2)^2 -q_1^2q_2^2$ and $R^{\mu\nu}=-g^{\mu\nu} +\frac{1}{X}\Big[ (q_1\!\cdot\!q_2)(q_1^{\mu}q_2^{\nu} +q_2^{\mu}q_1^{\nu})  -q_1^2 q_2^{\mu}q_2^{\nu} -q_2^2 q_1^{\mu}q_1^{\nu}  \Big]$, the form factor is defined as 
\begin{multline}
   \mathcal{M}^{\mu\nu\tau} = i\epsilon_{\rho\sigma\alpha}^{\;\;\;\;\;\,\tau}\Big[ R^{\mu\rho}R^{\nu\sigma}(q_1 -q_2)^{\alpha}\frac{q_1\!\cdot\!q_2}{m_A^2}F_A^{(0)}(q_1^2,q_2^2) 
   + R^{\nu\rho}\left(q_1^{\mu} -\frac{q_1^2}{q_1\!\cdot\!q_2} q_2^{\mu}\right)q_1^{\sigma}q_2^{\alpha}\frac{1}{m_A^2}F_A^{(1)}\left(q_1^2,q_2^2\right) \\
   + R^{\mu\rho}\left(q_2^{\nu} -\frac{q_2^2}{q_1\!\cdot\!q_2} q_1^{\nu}\right)q_2^{\sigma}q_1^{\alpha}\frac{1}{m_A^2}F_A^{(1)}\left(q_2^2,q_1^2\right)
   \Big].
\end{multline}
The outcome can be conveniently recast via the Schouten identities as
\begin{multline}
    \mathcal{M}^{\mu\nu\tau}=
    i\epsilon^{\nu q_1 \tau q_2}\left[q_1^{\mu} -q_2^{\mu}\frac{q_1^2}{q_1\!\cdot\!q_2}\right]\frac{1}{m_A^2}F_A^{(1)}
    +i\epsilon^{\mu q_2 \tau q_1}\left[q_2^{\nu} -q_1^{\nu}\frac{q_2^2}{q_1\!\cdot\!q_2}\right]\frac{1}{m_A^2}\bar{F}_A^{(1)}
    \\ i\epsilon^{\mu\nu q_1q_2}\frac{q_1\!\cdot\!q_2}{2 m_A^2 X} \left[ \bar{q}_{12}^{\tau}(q_1^2 -q_2^2) 
    -q_{12}^{\tau}\bar{q}_{12}^2 \right]F_A^{(0)} .
\end{multline}
The last piece, containing $q_{12}^{\tau}$, vanishes on-shell and the analogy to \cref{eq:tffJeger} is clear.\footnote{Note however that off-shell effects will be relevant for $(g-2)$ unless a transverse propagator is taken.} For $F_A^{(1)}$, the result is analogous to $F_A'$ up to the $(q_1\cdot q_2)$ overall term, that avoids kinematical zeroes. Finally, $F_A^{(0)}$ is, up to the additional {\textit{ad-hoc}} $q_{1,2}^2$-dependency induced, analogous to $F_A$ in \cref{eq:tffJeger}. Using \cref{eq:FredTrans} we obtain
\begin{align}
    C ={}& \frac{1}{m_A^2}\left[ \frac{(q_1\!\cdot\!q_2)^2 -q_2^2(q_1\!\cdot\!q_2)}{X}F_A^{(0)} -\frac{q_2^2}{q_1\!\cdot\!q_2} \bar{F}_A^{(1)} \right], &  
    B_2 ={}& \frac{1}{m_A^2}\left[ \bar{F}_A^{(1)} + \frac{q_1^2}{q_1\!\cdot\!q_2}F_A^{(1)} \right],  \\ 
    \bar{C} ={}&  \frac{1}{m_A^2}\left[ \frac{(q_1\!\cdot\!q_2)^2 -q_1^2(q_1\!\cdot\!q_2)}{X}F_A^{(0)} -\frac{q_1^2}{q_1\!\cdot\!q_2} F_A^{(1)} \right], & 
    \bar{B}_2 ={}& \frac{1}{m_A^2}\left[ F_A^{(1)} + \frac{q_2^2}{q_1\!\cdot\!q_2}\bar{F}_A^{(1)} \right].
\end{align}
Note however that such form factors have not been used so far to compute the contribution to $(g-2)$; instead, the ones in the previous subsection were employed~\cite{Pauk:2014rta}.

\subsection{\texorpdfstring{\boldmath{$\langle VVA \rangle$}}{<VVA>} basis}\label{app:VVA}

It can be shown that, up to overall factors, the axial meson contributions to the $\langle VVA \rangle$ Green's function corresponds to that of the axial meson transition form factors times an additional  $(1/i)\sqrt{2}F_A M_A(q_{12}^2 -M_A^2)^{-1}d^{abc}/2$ factor, that makes interesting to study the connection to the standard tensor basis for $\langle VVA \rangle$ that is employed in Refs.~\cite{Knecht:2003xy,Jegerlehner:2005fs,Kadavy:2016lys}\footnote{Ref.~\cite{Jegerlehner:2005fs} uses $\epsilon^{0123}=-1$ instead, that we adapt. Further, we omitted $i$ overall terms as they cancel in the transition from the axial form factors to the $\langle VVA \rangle$ function, and the overall $(8\pi^2)^{-1}$ in \cref{eq:vvabasis}.} 
\begin{align}
    \langle V_{\mu}(q_1)V_{\nu}(q_2)A_{\tau} \rangle ={}& \frac{\epsilon^{0123}}{8\pi^2}\Big\{ 
      -w_L\epsilon_{\mu\nu q_1 q_2}q_{12\tau} + w_T^{(+)}t_{\mu\nu\tau}^{(+)} 
      +w_T^{(-)}t_{\mu\nu\tau}^{(-)} +\widetilde{w}_T^{(-)}\tilde{t}_{\mu\nu\tau}^{(-)}
    \Big\}, \label{eq:vvabasis}\\
    t_{\mu\nu\tau}^{(+)}  ={}& 
    \epsilon_{q_1q_2\mu\tau}q_{1\nu} -\epsilon_{q_1q_2\nu\tau}q_{2\mu}
    -(q_1\!\cdot\!q_2)\epsilon_{\mu\nu\tau\bar{q}_{12}}
    -2\frac{(q_1\!\cdot\!q_2)}{q_{12}^2}\epsilon_{\mu\nu q_1q_2}q_{12\tau}, \\
    t_{\mu\nu\tau}^{(-)}  ={}& 
    \epsilon_{\mu\nu q_1q_2}\left[\bar{q}_{12\tau} -\frac{q_{12}\!\cdot\!\bar{q}_{12}}{q_{12}^2} q_{12\tau}\right], \\
    \tilde{t}_{\mu\nu\tau}^{(-)}  ={}& 
    \epsilon_{q_1q_2\mu\tau}q_{1\nu} +\epsilon_{q_1q_2\nu\tau}q_{2\mu}
    -(q_1\!\cdot\!q_2)[\epsilon_{\mu\nu\tau q_1} +\epsilon_{\mu\nu\tau q_2}].
\end{align}
Comparing to \cref{eq:GenAxTFF}, one can identify the form factors and recast them via the Schouten identities in terms of those in \cref{eq:axialFF}, showing that
\begin{align}
    C_A ={}& w_T^{(-)}\! +\widetilde{w}_T^{(-)}&
    B_{2A} ={}& \widetilde{w}_T^{(-)}&
    B_{2S} =& \!-\!w_T^{(+)}&
    C_S =& -\!w_L +\frac{q_1^2 +q_2^2}{q_{12}^2}w_T^{(+)}\! -\frac{q_{12}\!\cdot\!\bar{q}_{12}}{q_{12}^2} w_T^{(-)} \\
    w_T^{(-)}\!\! ={}& C_A -B_{2A}&  
    \widetilde{w}_T^{(-)}\!\! ={}& B_{2A}&  
    w_T^{(+)}\!\! =& \!-\!B_{2S}&  
    w_L =& -\![C_S +\frac{q_1^2+q_2^2}{q_{12}^2}B_{2S} + \frac{q_{12}\!\cdot\!\bar{q}_{12}}{q_{12}^2}(C_A\! -\!B_{2A})]. \nonumber
\end{align}
Indeed, the structure of the antisymmetric tensor formalism will guarantee a vanishing contribution of axial resonances to longitudinal degrees of freedom, this is, to $w_L$ above. This was ensured in our results since $C_A = B_{2A}$, and will persist in higher orders---see for instance the section below. Actually, this is crucial within the antisymmetric formalism since otherwise, in the chiral limit, it would spoil the anomaly that is fulfilled via different operators lacking intermediate axial mesons~\cite{Kadavy:2017nxc}. It is interesting to wonder how such an analogous result will appear in the HLbL in the OPE limit defined in~\cite{Melnikov:2003xd}, that connects the HLbL with the $\langle VVA \rangle$ correlator and fixes the longitudinal part. This is currently under investigation, but it is clear that, in the chiral limit, this cannot be attributed to heavy pseudoscalars~\cite{Colangelo:2019lpu} as it has been recently noted in Ref.~\cite{Cappiello:2019hwh}. 

\section{Higher orders \texorpdfstring{R\boldmath{$\chi$}T}{RChPT} estimation}
\label{app:HO}

As we showed, at LO in R$\chi$T, there is a single---and antisymmetric---form  factor, $B_{2A}=C_A$. In turn, the symmetric one(s), $B_{2S}$ (and $C_{S}$), despite their central role in $\gamma\gamma^*\to A$ transitions at low-energies, are relegated to higher orders in the chiral counting. In this section, we illustrate this assertion and the impact of the off-shell prescription.

Since a basis for the odd-parity sector in R$\chi$T within the antisymmetric tensor formalism contributing to the chiral LECs at $\mathcal{O}(p^{8}$) has not been completed yet, we select particular operators that should, in essence, capture the general features of higher order corrections contributing to $B_{2S}$:
\begin{equation}
\label{newoperators}    \kappa_X^{VV} \epsilon_{\mu\nu\rho\sigma}\langle V^{\mu\nu} \partial_{\alpha}V^{\alpha\rho} \partial_{\beta}A^{\beta\sigma} \rangle, \quad
    -\frac{\kappa_X^{\gamma\gamma}}{4} \epsilon_{\mu\nu\rho\sigma}\langle f_+^{\mu\nu} \partial_{\alpha}f_+^{\alpha\rho} \partial_{\beta}A^{\beta\sigma} \rangle,
\end{equation}
where, since we are interested in diagonal isospin elements, all (anti)commutators become trivial and thereby ignored. The analogous of \cref{eq:RChThaxialtoGG} becomes
\begin{equation}\label{eq:hoM}
    \mathcal{M}^{\mu\nu;\rho\lambda} = c_A q_{12}^{\rho} \left[
       \epsilon^{\mu\alpha\lambda q_1}(q_2^{\alpha}q_2^{\nu} -q_2^2g^{\nu\alpha})
      +\epsilon^{\nu\alpha\lambda q_2}(q_1^{\alpha}q_1^{\mu} -q_1^2g^{\mu\alpha})
    \right]F_{A\gamma^*\gamma^*}(q_1^2,q_2^2),
\end{equation}
where $c_A = \kappa_X^{\gamma\gamma(VV)}2\langle A \mathcal{Q}^2 \rangle = \kappa_X^{\gamma\gamma(VV)}\{ \frac{\sqrt{2}}{3}, \frac{5\sqrt{2}}{9}, \frac{2}{9} \}$ for $\{ a_1, f_1, f_1'\}$, is an isospin factor. Operators like the first one, will generate a $q_{1,2}^2$-dependent form factor. In particular
\begin{equation}
    F_{\{a_1,f_1\}\gamma^*\gamma^*} =  \frac{(\sqrt{2}F_{\rho\omega})^2 }{(q_1^2-M_{\rho\omega}^2)(q_2^2-M_{\rho\omega}^2)}, \quad
     F_{f'_1\gamma^*\gamma^*} =  \frac{(\sqrt{2}F_{\phi})^2 }{(q_1^2-M_{\phi}^2)(q_2^2-M_{\phi}^2)},
\end{equation}
while the second operator would produce a constant form factor. In general, there will be many more operators contributing, and several vector multiplets can be considered. Thereby, for simplicity, we will employ the result in \cref{eq:hoM} and append some generic form factor that we will fix from phenomenology. For that purpose, we compute the axial transition form factors by contracting $\mathcal{M}^{\mu\nu;\rho\lambda}$ in \cref{eq:hoM} with $\bra{0} A^{\rho\lambda} \ket{A}$ [see comments below \cref{eq:RChThaxialtoGG}]. We obtain ($\tilde{F}_{A\gamma^*\gamma^*}(0,0) =1$)
\begin{equation}
    B_{2S} = \frac{c_A}{m_A} q_{12}^2 \tilde{F}_{A\gamma^*\gamma^*}(q_1^2,q_2^2), \qquad 
    C_{S} = -\frac{q_1^2 +q_2^2}{q_{12}^2} B_{2S}.   
\end{equation}
Note that the relevant feature is the chiral $q_{12}^2\neq m_A^2$ suppression together with the nonvanishing value for $C_S$ that is key to avoid contributions to the longitudinal degrees of freedom, as anticipated. All these features already announce a chiral suppression if compared to the standard ones. In order to compute the $c_A$ coefficient above, we use the relation in \cref{eq:Awidth}, implying that 
\begin{equation}
    c_A = \pm\sqrt{ \frac{12\tilde{\Gamma}_{A\gamma\gamma}}{\pi\alpha^2 m_A^7}  } \longrightarrow   \tilde{c}_A\equiv m_A^3 c_A =  \pm\sqrt{ \frac{12\tilde{\Gamma}_{\gamma\gamma}}{\pi\alpha^2 m_A}  }.
\end{equation}
So far, we only have results for the $f_1$ and $f_1'$ resonances. Particularly, L3 Collaboration has found $\tilde{\Gamma}_{f_1\gamma\gamma} = 3.5(6)(5)$~keV~\cite{Achard:2001uu} and $\tilde{\Gamma}_{f'_1\gamma\gamma} = 3.2(6)(7) $~keV~\cite{Achard:2007hm}, respectively. These imply that $|\tilde{c}_{f_1}| = 0.44(5)$ and  $|\tilde{c}_{f'_1}| = 0.40(6)$; finally, we make a generous estimate $\tilde{\Gamma}_{a_1\gamma\gamma}\in (1,3)$~keV (see also \cite{Leutgeb:2019gbz}), so that  $\tilde{c}_{a_1\gamma\gamma} \in (0.2,0.4)$. Moreover, the experimental results also extract dipole masses $\Lambda_{f_1} = 1.04(6)(5)$~GeV and $\Lambda_{f'_1} = 0.926(72)(32)$~GeV, respectively (we will assume $\Lambda_{a_1} = 1.0(1)$~GeV). We will use thereby
\begin{equation}
    \mathcal{M}^{\mu\nu;\rho\lambda} = q_{12}^{\rho} \left[
       \epsilon^{\mu\alpha\lambda q_1}(q_2^{\alpha}q_2^{\nu} -q_2^2g^{\nu\alpha})
      +\epsilon^{\nu\alpha\lambda q_2}(q_1^{\alpha}q_1^{\mu} -q_1^2g^{\mu\alpha})
    \right]\frac{\tilde{c}_A}{m_A^3}\frac{\Lambda_A^8}{(q_1^2 -\Lambda_A^2 )^2(q_2^2 -\Lambda_A^2 )^2 }
\end{equation}
as the input for the respective contributions to \amu{}.

Computing only this contribution to \amu{} we find, in units of $10^{-11}$, $\Delta a_{\mu}^{a_1} = -0.2(1)$, $\Delta a_{\mu}^{f_1} = -0.42$ and  $\Delta a_{\mu}^{f_1'} = -0.06$. Adding this contribution to that in \cref{eq:RChThaxialtoGG} produces interference terms as well. Depending on the relative sign we find $\Delta a_{\mu}^{a_1} = \mp0.03(1)$, $\Delta a_{\mu}^{f_1} = \mp 0.07$ and $\Delta a_{\mu}^{f_1'} = \mp 0.007$. These corrections are similar in size to those discussed in \cref{sec:Contributiontoamu}---that is to be expected since, effectively, they are of the same order. Still, these are much smaller than what estimated in Ref.~\cite{Pauk:2014rta}, despite on-shell our results are equivalent. Once more, this shows the potential systematic uncertainties of existing approaches even if experimental input is used.

\section{Phenomenological information on the relevant parameters of the \texorpdfstring{R\boldmath{$\chi$}T}{RChPT} Lagrangian}
\label{app:phenoRChT}

For the spin-one meson nonets, in application of the large-$N_C$ limit~\cite{tHooft:1973alw}, we have considered ideal mixing between the isoscalar component of the octet and the additional isosinglet state completing the nonet. This way, we will have the following diagonal elements of the nonets in flavor space:
\begin{equation}
 (V_{11},V_{22},V_{33})^{\mu\nu}\,=\,\left(\frac{\rho^0+\omega}{\sqrt{2}},\frac{-\rho^0+\omega}{\sqrt{2}},\phi\right)^{\mu\nu}\,,\quad (A_{11},A_{22},A_{33})^{\mu\nu}\,=\,\left(\frac{a_1^0+f_1}{\sqrt{2}},\frac{-a_1^0+f_1}{\sqrt{2}},f_1'\right)^{\mu\nu}\,,
\end{equation}
where $f_1\sim f_1(1285)$ and $f_1'\sim f_1(1420)$. The leading breaking of the $U(3)$ symmetry splits the heaviest components of each nonet ($\phi$ and $f_1'$) from its partners. In the large-$N_C$ and isospin symmetry limits, the Lagrangian bilinear in the spin-one fields of the same type (either $VV$ or $AA$) in the even-intrinsic parity sector \cite{Cirigliano:2006hb} produces the mass splittings \cite{Cirigliano:2003yq, Guo:2014yva} ($M_V$ and $M_A$ are the large-$N_C$ masses of the whole nonet before the symmetry breaking, which is induced by nonvanishing $e^V_m$)
\begin{eqnarray} \label{masseswithU(3)breaking}
 M_\rho^2=M_V^2-4e^V_m m_\pi^2=M_\omega^2\,,\quad M_\phi^2=M_V^2-4e^V_m (2m_K^2-m_\pi^2)\,,\nonumber\\
 M_{a_1}^2=M_A^2-4e^A_m m_\pi^2=M_{f_1}^2\,,\quad M_{f_1^\prime}^2=M_A^2-4e^A_m (2m_K^2-m_\pi^2)\,.
\end{eqnarray}
 From the best fit in Ref. \cite{Guevara:2018rhj} one has $M_V=(791\pm6)$ MeV and $e^V_m=-0.36\pm0.10$, which deviates clearly from the fit to mass spectrum that is obtained if one identifies the states in the large-$N_C$ limit with the physical states, yielding $M_V\sim764.3$ MeV and $e^V_m\sim-0.28$ \cite{Guo:2014yva}. It is well understood, however, that such departures occur \cite{Masjuan:2007ay}. In absence of data on the axial-vector transition form factors that could help us to verify in which way $M_A$ and $e^A_m$ differ from the naive values that are obtained fitting the axial-vector meson nonet with the above formulas, and as $M_V^2$ and $M_A^2$ are connected by short-distance constraints \cite{Weinberg:1967kj}, we will assume that the shift induced is analogous to the one for the vector mesons. In this way, we obtain $M_A=(1310\pm44)$ MeV and $e^A_m=-0.35\pm0.13$, where the conservative error is estimated so as to include the naive values of $M_A$ and $e^A_m$ at one standard deviation. According to the preceding discussion, we will use in the following
 \begin{equation}\label{largeNCmasses}
  M_V=(791\pm6)\, \mathrm{MeV}\,,\;e^V_m=-0.36\pm0.10\,,\;M_A=(1310\pm44)\, \mathrm{MeV}\,,\;e^A_m=-0.35\pm0.13\,,
 \end{equation}
 so that, in this limit, the common mass for the isotriplet and isoscalar states of the spin-one octets is
 \begin{equation}\label{largeNCmasses8}
  M_{\rho\omega}=(808\pm8)\, \mathrm{MeV}\,,\quad M_{a_1f_1}=(1320\pm44)\,\mathrm{MeV}\,,
 \end{equation}
 and the common mass for the extra isoscalar state is
 \begin{equation}\label{largeNCmasses1}
  M_\phi=(1144\pm80)\,\mathrm{MeV}\,,\quad M_{f_1^\prime}=(1543\pm96)\,\mathrm{MeV}\,.
 \end{equation}
 
 We observe that $M_{\rho\omega}$ is $\sim 4\%$ larger than its experimental (isospin-averaged) value, while $M_\phi$ is $\sim12\%$ larger than its measurement. We will assume a similar deviation for the corresponding states in excited multiplets.
 
The considered flavor-symmetry breaking also affects the coupling of the vector meson resonances to the photon (encoded in the $F_V$ couplings). However, as shown in Ref. \cite{Guevara:2018rhj}, the corresponding leading shifts are given in terms of a single coupling ($\lambda_6^V$ in Ref. \cite{Cirigliano:2006hb}), that vanishes according to short-distance QCD constraints \cite{Guevara:2018rhj}. Thus, $F_\rho\sim F_\omega\sim F_\phi\sim F_V$, within our setting (and similarly for the excited vector resonances). Since $F_{a_1}\neq F_{f_1}\neq F_{f_1'}$ is induced  in complete analogy, we will take the coupling of the axial-vector resonances to the axial current ($F_A$) in the $U(3)$ symmetry limit, as their breaking given by $\lambda_6^A$ vanishes by asymptotic conditions \cite{Guevara:2018rhj}.
 
 As noted before, the axial-vector contribution to the hadronic light-by-light piece of the muon anomalous magnetic moment, within R$\chi$T, only depends on the product of couplings $F_{V_i} \kappa_5^{{V_i}A}$ for the different $i=1,2,3...$ vector multiplets and on the (axial-)vector-resonance masses. Moreover, the high-energy behavior of our form factor links additional $F_{V_i} \kappa_5^{{V_i}A}$ factors to $F_{V_1} \kappa_5^{{V_1}A}$, while the masses of the corresponding multiplets are needed inputs in this case, see \cref{relations3V}. In order to determine $F_{V_1} \kappa_5^{{V_1}A}$, we follow Refs.~\cite{Kadavy:2016lys,Kadavy:2017nxc}, where the OPE condition for the $VVA$ Green's function up to $\mathcal{O}(1/p^4)$ demands---when matching the R$\chi$T result to it---that
 \begin{equation}
 \label{OPEprediction2}
  \kappa_5^{VA}=\kappa_3^{VV}\frac{F_V}{F_A} 
  = -\frac{N_CM_V^2}{64\pi^2F_VF_A} \rightarrow 
  F_V\kappa_5^{VA}=-\frac{N_CM_V^2}{64\pi^2F_A}\,,
 \end{equation}
where the second equality follows from the constraint for $\kappa_3^{VV}$ in Ref. \cite{Roig:2013baa} and $F_A\in[130,150]$ MeV \cite{Dumm:2009va, Nugent:2013hxa}. We note that, with only one vector and one axial-vector multiplet, the first Weinberg rule is $F_V^2-F_A^2=F^2$ that, using $F_V=\sqrt{3}F\sim160$ MeV \cite{Roig:2013baa} (which is quite well satisfied phenomenologically \cite{Dumm:2009va, Nugent:2013hxa,  Shekhovtsova:2012ra}), yields $F_A=\sqrt{2}F\sim130$ MeV. Employing the previous values in \cref{OPEprediction2}, one finds $F_V\kappa_5^{VA}\sim(-21.3\pm1.5)$ MeV, 
with reasonably little uncertainty and that we shall employ in our calculations. 

Instead, one could use $A\to V\gamma$ decays, whose amplitude reads
\begin{equation}
    \Gamma(A\to V\gamma) = \frac{2}{3}\alpha |\kappa_5^{VA}|^2 m_A\left( 1- \frac{m_V^2}{m_A^2} \right)^3\left( 1+ \frac{m_A^2}{m_V^2}\right)[\operatorname{tr}(\{V,A\} Q)]^2.
\end{equation}
Employing the $f_1(1285)\to\rho\gamma$ branching fraction \cite{Tanabashi:2018oca} we find $|\kappa_5^{VA}|=0.45\pm0.06$. However, a recent measurement by CLAS Collaboration~\cite{Dickson:2016gwc} implies a much smaller width, that would imply $|\kappa_5^{VA}|=0.27\pm0.06$, much closer to the value $\kappa_5^{VA}=-0.12\pm0.02$ obtained using short-distance constraints \cite{Ecker:1988te, Dumm:2009va, Roig:2013baa, Nugent:2013hxa} or phenomenological determinations of $F_V$ and $F_A$ in  \cref{OPEprediction2}. Further, given the $\rho$-meson width, additional operators involving pion fields might be relevant as well. For these reasons, we advocate to adopt the value implied by the short-distance constraints and emphasize the need for future measurements.

The last input to be fixed are the masses of the vector meson excitations. This will be done using the corresponding generalization of \cref{masseswithU(3)breaking} and assuming that $M_{\rho'\omega'}$ and $M_{\phi'}$ exceed their (isospin-averaged) PDG values by $\sim4\%$ and $\sim12\%$, respectively (as it happens with the lightest vector multiplet), and analogously with $M_{\rho''\omega''}$ and $M_{\phi''}$. In this way, we estimate
\begin{eqnarray}
    M_{\rho'\omega'}\,=\,(1.51\pm0.03)\,\mathrm{GeV}\,,\;M_{\phi'}\,=\,(1.88\pm0.03)\,\mathrm{GeV}\,,\nonumber\\
    M_{\rho''\omega''}\,=\,(1.78\pm0.03)\,\mathrm{GeV}\,,\;M_{\phi''}\,=\,(2.45\pm0.03)\,\mathrm{GeV}\,.
\end{eqnarray}

\section{Functions involved in the HLbL computation}\label{sec:appHLBL}

The axial-meson contribution to the HLbL tensor can be succinctly expressed---in an obvious gauge invariant way---as follows:
\begin{equation}
  \Pi^{\textrm{HLbL};A}_{\mu\nu\rho\sigma}\varepsilon_1^{\mu}\varepsilon_2^{\nu}\varepsilon_3^{\rho}\varepsilon_4^{\sigma}   = -4iF_A(q_1^2,q_2^2)F_A(q_3^2,q_4^2)\Delta_F^R(q_{12})^{\alpha\beta,\bar{\alpha}\bar{\beta}} (\tilde{F}_2 F_1)_{\alpha\beta}(\tilde{F}_4 F_3)_{\bar{\alpha}\bar{\beta}} + (2\leftrightarrow3) +(2\leftrightarrow4) , \label{eq:HLBLshort}
\end{equation}
The functions $K_i(Q_1^2,Q_2^2,t)$ introduced in \cref{eq:lblmaster}, arising in the $a_{\mu}^{\textrm{HLbL};A}$ evaluation, are given by
\begin{multline}
K_1(Q_1^2,Q_2^2,t)=\frac{8 F_A(Q_1^2,Q_3^2)F_A(Q_2^2,0)}{(Q_2^2 +m_A^2)}
 \Big[  \frac{2m_{\mu}^2}{Q_1^2} \left(\frac{Q_1 \left(t^2-1\right) (2 Q_1+Q_2 t)}{m_A^2}-\frac{2 Q_1^2+3 Q_1 Q_2 t+Q_2^2}{Q_2^2}\right)      \\  +(1-R_{m_1}) \left(\frac{2 Q_1^2 \left(t^2-1\right)+Q_1 Q_2 t \left(t^2-5\right)+2 Q_2^2 \left(t^2-3\right)}{m_A^2}-\frac{2 Q_1^2+7 Q_1 Q_2 t+6 Q_2^2}{Q_2^2}\right)  \\  -\frac{(1-R_{m_2})}{m_A^2 Q_1^2} \left(m_A^2 \left(6 Q_1^2+8 Q_1 Q_2 t+Q_2^2\right)+4 Q_1 Q_2^2 (Q_1+Q_2 t)\right)  \\  + 4 X \left(-\frac{Q_2^2 \left(2 Q_3^2+Q_2^2 \left(1-t^2\right)\right)}{m_A^2}-3 Q_3^2+m_{\mu}^2 \left(\frac{2 Q_2 t}{Q_1}+\frac{4 Q_1 t}{Q_2}+4 t^2+2\right)\right) \Big], \label{eq:K1}
\end{multline}
\begin{multline}
 K_2(Q_1^2,Q_2^2,t)=\frac{4 F_A(Q_1^2,Q_2^2)F_A(Q_3^2,0)}{(Q_3^2 +m_A^2)} \Big[
 \frac{4 \left(Q_1^2-Q_2^2\right) X \left(Q_1 Q_2 Q_3^2+2 m_A^2 m_{\mu}^2 t\right)}{m_A^2 Q_1 Q_2} \\ 
 +\frac{(1-R_{m_1}) \left(m_A^2 Q_1 (Q_2 t-Q_1)+Q_2 \left(2 Q_1^3 t+Q_1^2 Q_2 \left(3 t^2-1\right)+Q_1 Q_2^2 t \left(t^2-3\right)-2 Q_2^3\right)\right)}{m_A^2 Q_2^2} \\ 
 -\frac{(1-R_{m_2}) \left(m_A^2 Q_2 (Q_1 t-Q_2)+Q_1 \left(-2 Q_1^3+Q_1^2 Q_2 t \left(t^2-3\right)+Q_1 Q_2^2 \left(3 t^2-1\right)+2 Q_2^3 t\right)\right)}{m_A^2 Q_1^2} \\ 
 -\frac{2 m_{\mu}^2 \left(Q_1^2-Q_2^2\right) \left(m_A^2+Q_1 Q_2 t \left(t^2-1\right)\right)}{m_A^2 Q_1^2 Q_2^2} \Big].  \label{eq:K2}
\end{multline}
where the following functions, together with $Q_3^2 = Q_1^2 +Q_2^2 +2Q_1Q_2 t$, have been employed 
\begin{equation}
    R_{m_i} = \sqrt{1+\frac{4m^2}{Q_i^{2}}}, \ \
    z = \frac{Q_1Q_2}{4m_{\mu}^2}(1-R_{m_1})(1-R_{m_2}), \ \
    X = \frac{(1-t^2)^{-1/2}}{Q_1 Q_2} \arctan\left(\frac{z\sqrt{1-t^2}}{1-zt} \right).
\end{equation}

It is also interesting to discuss the asymptotic behavior of the integrands. From the definition in \cref{eq:lblmaster}, and for constant $F_A(Q_1^2,Q_2^2)\to 1$ form factors we obtain for $w_1$
\begin{align}
    \lim_{Q\to\infty} \int_{-1}^1 dt \ w_1(Q,Q,t) ={}& \frac{70\pi^2 m_{\mu}^2}{3m_A^2}, \\
    \lim_{Q_1\to\infty} \int_{-1}^1 dt \ w_1(Q_1,Q_2,t) ={}& \frac{8\pi^2 Q_2^3}{3m_A^2Q_1(m_A^2+Q_2^2)}
        \Big[ 3(Q_2^2 -m_{\mu}^2) \nonumber\\ {}& \qquad +R_{m_2}(6m_A^2 +4Q_2^2) + \frac{Q_2^2}{2m^2}(6m_A^2 +7Q_2^2)(1-R_{m_2})  \Big],\\
    \lim_{Q_2\to\infty} \int_{-1}^1 dt \ w_1(Q_1,Q_2,t) ={}& \frac{2\pi^2 Q_1^3}{9m_A^2 Q_2} 
         \Big[ 68 R_{m_1} -42 +\frac{13 Q_1^2}{m^2}(1 -R_{m_1})
         \Big];
\end{align}
while for the second case, $w_2$, we find
\begin{align}
    \lim_{Q\to\infty} \int_{-1}^1 dt \ w_2(Q,Q,t) ={}& 0, \\
    \lim_{Q_{1(2)}\to\infty} \int_{-1}^1 dt \ w_2(Q_1,Q_2,t) ={}& \pm \frac{\pi^2 Q_{2(1)}^3}{3m_A^2Q_{1(2)}}
        \Big[ 14 -8R_{m_{2(1)}} +\frac{3Q_{2(1)}^2}{m_{\mu}^2}(1 -R_{m_{2(1)}})  \Big], 
\end{align}
the first result and the relations among the large-$Q_{1,2}$ limits due to the antisymmetric properties of the integrand.
Clearly, the asymptotic results set constraints on the form factor asymptotic behavior. In particular, we find that the large $Q_{1(2)}$ limits require the form factors to fall, at least, as $Q_{1(2)}^{-1}$ that, due to the antisymmetric nature of the form factor, demands at least a dipole form.

\section{Operator product expansion}\label{app:ope}

For two highly virtual photons, $q_{1,2} \simeq \pm \lambda q$, with $\lambda\to\infty$, so that $q_1+q_2 = \mathcal{O}(1)$, while $q_1-q_2 = \mathcal{O}(\lambda)$, one can use the operator product expansion, which is valid for large spacelike momenta. As a result, one finds~\cite{Melnikov:2003xd}
\begin{align}
    (2\pi)^4 \delta^{(4)}(q_1+q_2-q_A)\mathcal{M}^{\mu\nu\tau} \varepsilon_{A\tau} ={}& 
    i\int d^4xd^4y~e^{iq_1\cdot x}e^{iq_2\cdot y} \bra{0} T\{ j^{\mu}(x) j^{\nu}(y) \} \ket{A} \\
    ={}& \frac{-2i}{\hat{q}^2}\epsilon^{\mu\nu\rho\hat{q}}\int d^4z~e^{i(q_1+q_2)\cdot z} \bra{0} j_{5\rho}(z) \ket{A},
\end{align}
where $\hat{q}=(q_1-q_2)/2$ and $j_{5\rho}= \bar{q}\gamma_{\rho}\gamma^5 \hat{Q}^2 q$, with $\hat{Q}$ the charge operator. This implies, adopting $\bra{0} j_{5\rho} \ket{A} \equiv \sqrt{2}F_A m_A \epsilon_{A\rho}\operatorname{tr}(\hat{Q}^2 A)$, that
\begin{equation}
    \mathcal{M}^{\mu\nu\tau}\epsilon_{A\tau} \to -\frac{4i}{(q_1 -q_2)^2} \sqrt{2}F_A m_A \operatorname{tr}(\hat{Q}^2 A)
    \epsilon^{\mu\nu\varepsilon_A \bar{q}_{12}} = 
    -\frac{i}{\hat{q}^2} \sqrt{2}F_A m_A \operatorname{tr}(\hat{Q}^2 A)
    \epsilon^{\mu\nu\varepsilon_A \bar{q}_{12}}. 
\end{equation}
Comparing to \cref{eq:axialFF}, the former puts the following constraint
\begin{equation}\label{eq:OPEcons}
    \lim_{Q^2\to\infty} = B_{2S}(-Q^2,-Q^2) = \frac{\sqrt{2}F_A m_A}{Q^4} + \mathcal{O}(Q^{-6}). 
\end{equation}
while no restrictions arise for $B_{2A}, C_{A,S}$ form factors. Further, the OPE cannot be used to set constraints on the singly virtual TFFs. However, at this respect, experimental data seems to favor, for all single-virtual form factors, a $Q^{-4}$ high-energy scaling as well.

\bibliographystyle{apsrev4-1}
\bibliography{references}

\end{document}